\documentclass[conference]{IEEEtran}
\IEEEoverridecommandlockouts
\pdfoutput=1
\usepackage{cite}
\usepackage{amsmath,amssymb,amsfonts}
\usepackage{algorithmic}
\usepackage{textcomp}
\usepackage{xcolor}

\usepackage{url}

\usepackage[ruled,vlined,linesnumbered]{algorithm2e}
\usepackage{multirow}
\usepackage{booktabs}
\usepackage{subfigure}
\usepackage[justification=centering]{caption}
\usepackage[pdftex]{graphicx}

\def\BibTeX{{\rm B\kern-.05em{\sc i\kern-.025em b}\kern-.08em
    T\kern-.1667em\lower.7ex\hbox{E}\kern-.125emX}}
\begin{document}

\title{HyperMAN: Hypergraph-enhanced Meta-learning Adaptive Network for Next POI Recommendation}

\author{
\IEEEauthorblockN{Jinze Wang\textsuperscript{1,$\dag$}, Tiehua Zhang\textsuperscript{2,$\dag$}, Lu Zhang\textsuperscript{3,*}, Yang Bai\textsuperscript{3}, Xin Li\textsuperscript{4}, Jiong Jin\textsuperscript{1}}
\IEEEauthorblockA{\textsuperscript{1}School of Engineering, Swinburne University of Technology, Australia}
\IEEEauthorblockA{\textsuperscript{2}School of Computer Science and Technology, Tongji University, China
}
\IEEEauthorblockA{\textsuperscript{3}School of Cybersecurity, Chengdu University of Information
Technology, China
}
\IEEEauthorblockA{\textsuperscript{4}YSTC Energy Pty Ltd, Australia
}
\IEEEauthorblockA{\{jinzewang, jiongjin\}@swin.edu.au \{tiehuazhang\}@tongji.edu.cn\\ \{zhang\_lu010\}@outlook.com \{Alicepub\}@163.com \{lixin\}@ystcenergy.com} 
\thanks{$^\dag$ Both authors contributed equally to this work.}
\thanks{$^*$ Corresponding author: Lu Zhang}
\thanks{This work was partially supported by the National Natural Science Foundation of China under Grant 62372330, the Start-up Research Grant of Chengdu University of Information Technology under Grant KYTZ2023038, and the Natural Science Foundation of Sichuan Province under Grant 2024NSFSC1478.}}

\maketitle

\begin{abstract}
Next Point-of-Interest (POI) recommendation aims to predict users' next locations by leveraging historical check-in sequences. 
Although existing methods have shown promising results, they often struggle to capture complex high-order relationships and effectively adapt to \textcolor{black}{diverse user behaviors}, particularly when addressing the \textcolor{black}{cold-start issue}. 
To address these challenges, we propose Hypergraph-enhanced Meta-learning Adaptive Network (HyperMAN), a novel framework that integrates heterogeneous hypergraph modeling with a difficulty-aware meta-learning mechanism for next POI recommendation. Specifically, three types of heterogeneous hyperedges are designed to capture high-order relationships: user visit behaviors at specific times (Temporal behavioral hyperedge), spatial correlations among POIs (spatial functional hyperedge), and user long-term preferences (user preference hyperedge). Furthermore, a diversity-aware meta-learning mechanism is introduced to dynamically adjust learning strategies, considering users behavioral diversity. Extensive experiments on real-world datasets demonstrate that HyperMAN achieves superior performance, effectively addressing cold start challenges and significantly enhancing recommendation accuracy.
\end{abstract}


\section{Introduction}
Next point-of-interest (POI) recommender systems have gained significant attention to help users identify potential interesting locations, thereby mitigating information overload and supporting location-based industries such as travel planning and advertising~\cite{liu2023mandari,wang2023meta}. 
As users' activities are often restricted to a specific city, most research focuses on city-level check-ins and develops next POI recommenders. 
However, this focus is often hindered by the cold-start problem, since a lack of sufficient historical data for users or locations significantly impairs the recommendation performance.

Two primary paradigms have emerged to address such a challenge: sequential-based and graph-based methods. 
Sequential-based methods focus on modeling transitional patterns, evolving from traditional techniques like Markov Chains~\cite{cheng2013you} to advanced deep learning architectures such as recurrent neural networks (RNNs)~\cite{zhao2018personalized} and transformers~\cite{halder2021transformer}. 
Despite their strengths, these methods emphasize capturing sequential dependencies while often overlooking other critical relational aspects, e.g., spatial correlations among POIs. 
Motivated by the remarkable success of graph neural networks (GNNs), graph-based methods have been introduced to next POI recommendation, aiming to effectively capture complex structural relationships among locations\textcolor{black}{~\cite{wang2023adaptive, lim2020stp}}. 
However, these methods overlook the higher-order relations beyond pairwise connections. Drawing inspiration from the ability of hypergraphs to represent higher-order relationships, recent research has utilized hypergraph structures to model users' historical trajectories\textcolor{black}{~\cite{tan2024heterogeneous,wang2023eedn, lai2024disentangled}}, achieving considerable improvements in next POI recommendation. 
Although such hypergraph-based approaches have achieved state-of-the-art performance in next POI recommendation, two critical issues remain under-explored.

\begin{figure}[t]
\centering
\includegraphics[width=0.43\textwidth]{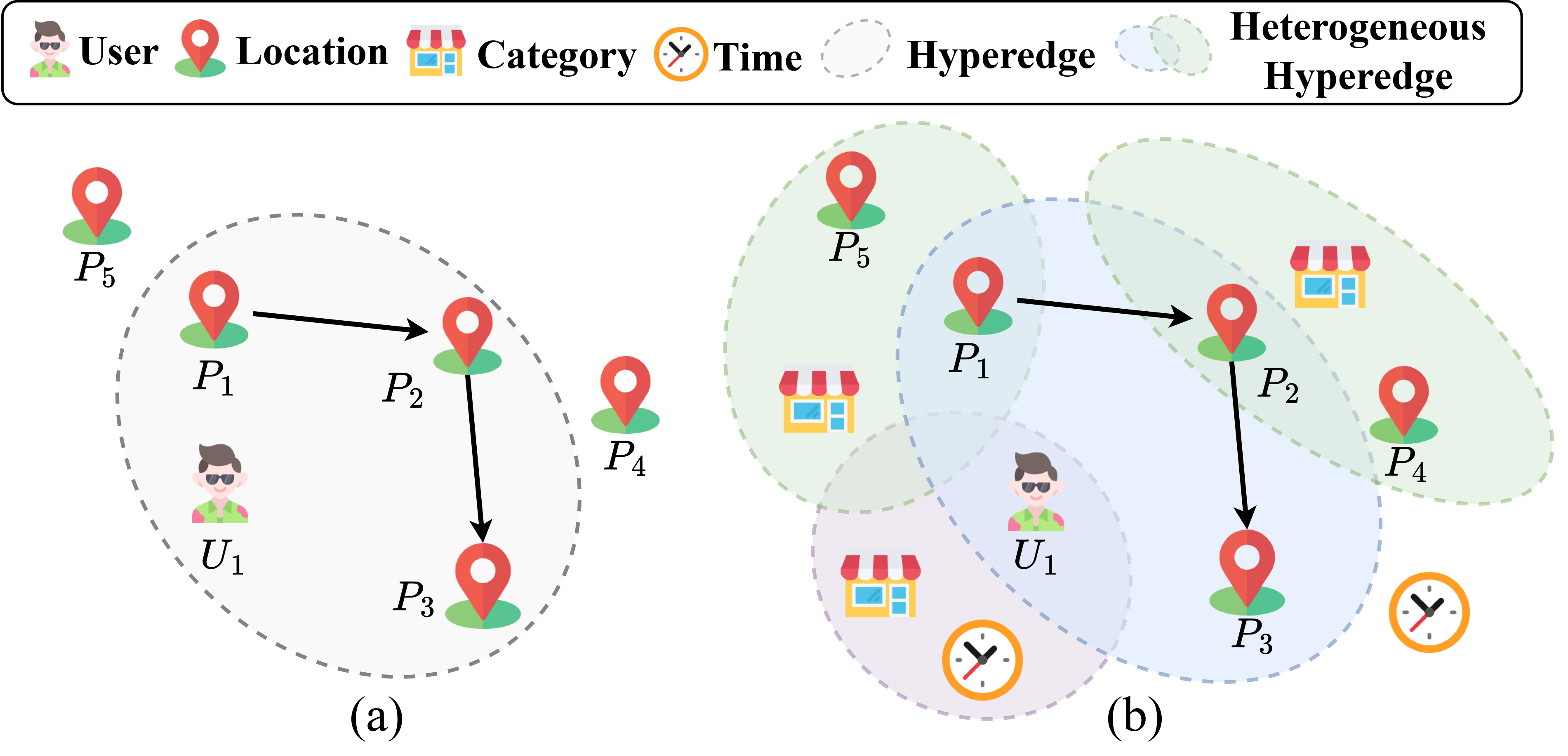}
\caption{Comparison between (a) a hypergraph and (b) a heterogeneous hypergraph.
}\label{fig:example}
\vspace{-0.3in}
\end{figure}

\textbf{Heterogeneous  Relationship.} 
In the context of next POI recommendation, interactions among users, POIs, categories, and time involve heterogeneous relationships. 
Existing studies overlook the fact that the interactions between users and POIs often exhibit heterogeneous properties. 
For example, when a user visits an office, restaurant, and shopping mall, the connections between location nodes and their respective categories (e.g., work, dining, shopping) can be represented through different types of heterogeneous hyperedges. Fig.~\ref{fig:example} shows the difference between hypergraphs and heterogeneous hypergraphs. This distinction enables the model to differentiate between the various types of locations and user behaviors. 

\textbf{Behavioral Diversity.} \textcolor{black}{Heterogeneous hypergraphs excel at capturing rich information, they may struggle to effectively handle unseen relationships associated with users, making it challenging to achieve comprehensive and high-quality representations when addressing cold-start problems~\cite{cai2023user}.} 
Motivated by the success of meta-learning in learning valuable patterns from limited user-item interactions, existing work takes advantage of meta-learning approaches to address the issue of cold start~\cite{wang2023adaptive,wang2023meta}. 
In particular, some studies integrate region-level user preferences through knowledge transfer or by simultaneously capturing user preferences and knowledge associations~\cite{sun2021mfnp,du2022metakg}.
\textcolor{black}{However, these approaches employ a fixed learning rate for all meta-learning tasks when addressing the cold-start problem, failing to account for the diversity in behavioral patterns across different users. 
For example, for users with regular, simple routines (e.g., commuting between home and office), the fixed learning rate leads to excessive adaptation that recommends unnecessary locations. 
While for users showing diverse visiting behaviors, the same learning rate results in insufficient adaptation that fails to capture their varying preferences. 
This ``one-size-fits-all" learning strategy significantly impairs the model's ability to handle cold-start users with different behavioral patterns.} 

\begin{figure}[t] 
\footnotesize
\centering
\includegraphics[width=0.4\textwidth]{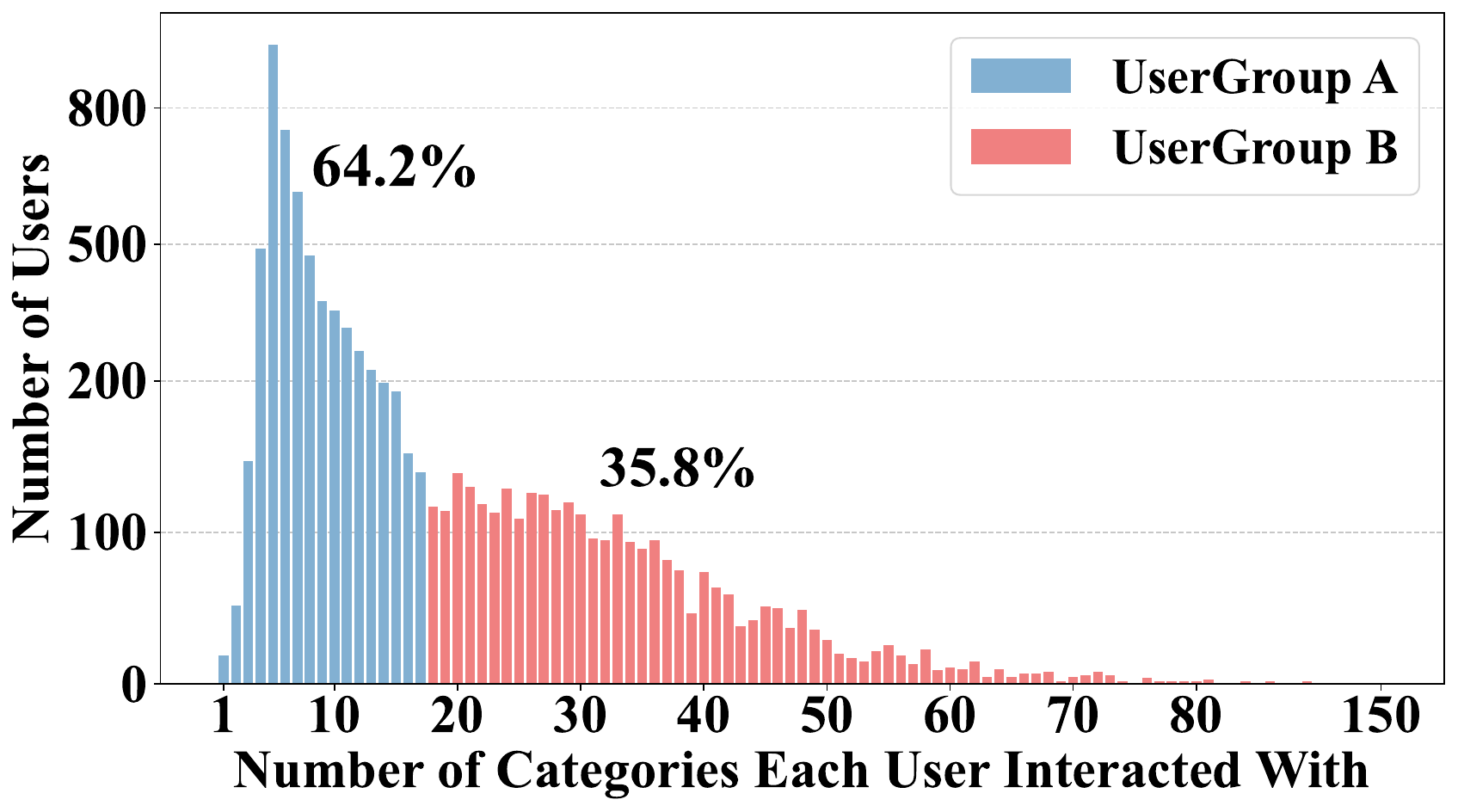} 
\caption{Comparison behavioral diversity.}
\label{fig:most_cate_vis}
\vspace{-0.2in}
\end{figure}

Fig.~\ref{fig:most_cate_vis} compares user behavioral diversity between two groups. Users are categorized based on the number of POI categories they interact with: Group A (regular and simple behaviors) and Group B (diverse behaviors). Group A, comprising 64.2\% of users, typically visits 1-15 categories, with 18\% focusing on five categories. In contrast, Group B (35.8\%) exhibits significantly broader interests, covering 20-80 categories, with some exploring up to 150. This diversity affects next POI recommendations—simpler behaviors allow accurate predictions with limited data, while diverse behaviors require higher-order relationships to capture complex preferences and mitigate cold-start issues.

Accordingly, we propose a novel Hypergraph-enhanced Meta-learning Adaptive Network (HyperMAN), which combines heterogeneous hypergraph modeling with a diversity-aware meta-learning mechanism for next POI recommendation. 
\textcolor{black}{Specifically, HyperMAN incorporates three types of heterogeneous hyperedges to capture different aspects of high-order relationships}: \textit{temporal behavioral hyperedges} representing user visit patterns at specific times, \textit{spatial functional hyperedges} \textcolor{black}{capturing relationships between nearby POIs of the same category}, and \textit{user preference hyperedges} reflecting user long-term preferences. Additionally, a \textit{diversity-aware meta-learning} mechanism is introduced to dynamically adapt learning strategies by accounting for \textcolor{black}{user behavioral diversity, which is measuring how user total visits are spread across different POI categories}. 

In sum, our main contributions are as follows: 
\begin{enumerate}
    \item \textcolor{black}{To the best of our knowledge, we are the first to explore heterogeneous hypergraph with meta-learning to address the issue of the cold-start in next POI recommendation.}
    \item We construct three types of heterogeneous hyperedges to capture complex high-order relationships, \textcolor{black}{including temporal behavioral hyperedges, spatial functional hyperedges, and user preference hyperedges.}
    \item We design a meta-learning training mechanism that accounts for \textcolor{black}{user behavioral diversity, dynamically adjusting learning rates to address the adaptation challenges posed by varying user behaviors.} 
    \item We conduct extensive experiments on four datasets to validate the superiority of HyperMAN over state-of-the-art methods.
\end{enumerate}

\section{Related Work}

\noindent\textbf{Graph and Hypergraph for Next POI Recommendation.}
As spatial correlations between POIs are essential for the next POI recommendation task, recent GNN-based methods are proposed to capture such correlations for improved recommendations.
For example, STP-UDGAT~\cite{lim2020stp} consists of three types of POI relationship graphs by considering user preferences, temporal aspects, and geographical relationships. 
KBGNN~\cite{ju2022kernel} constructs a geographical graph and uses a message-passing neural network to capture the topological geographical influences. 
These methods address the cold-start problem to some extent, however, they primarily focus on pairwise relationships and neglect high-order interactions among users and POIs. 
Subsequently, hypergraph-based methods are devised to capture high-order relationships~\cite{zhang2024dshgt, zhang2025learning}. 
DCHL~\cite{lai2024disentangled} designs three hypergraphs that incorporate collaborative, transitional, and geographical information. 
STHGCN~\cite{yan2023spatio} utilizes a hypergraph to capture trajectory-level information and collaborative signals from other users.
However, they overlook the fact that interactions between users and POIs often exhibit heterogeneous relationships. 
Hence, incorporating the diversity of interactions through heterogeneous hypergraphs could potentially improve model performance.
However, these methods overlook the fact that interactions between users and POIs often involve heterogeneous relationships. Incorporating the diversity of these interactions through heterogeneous hypergraphs could significantly enhance model performance.

\begin{figure*}[ht]
\centering 
\includegraphics[width=0.75\textwidth]{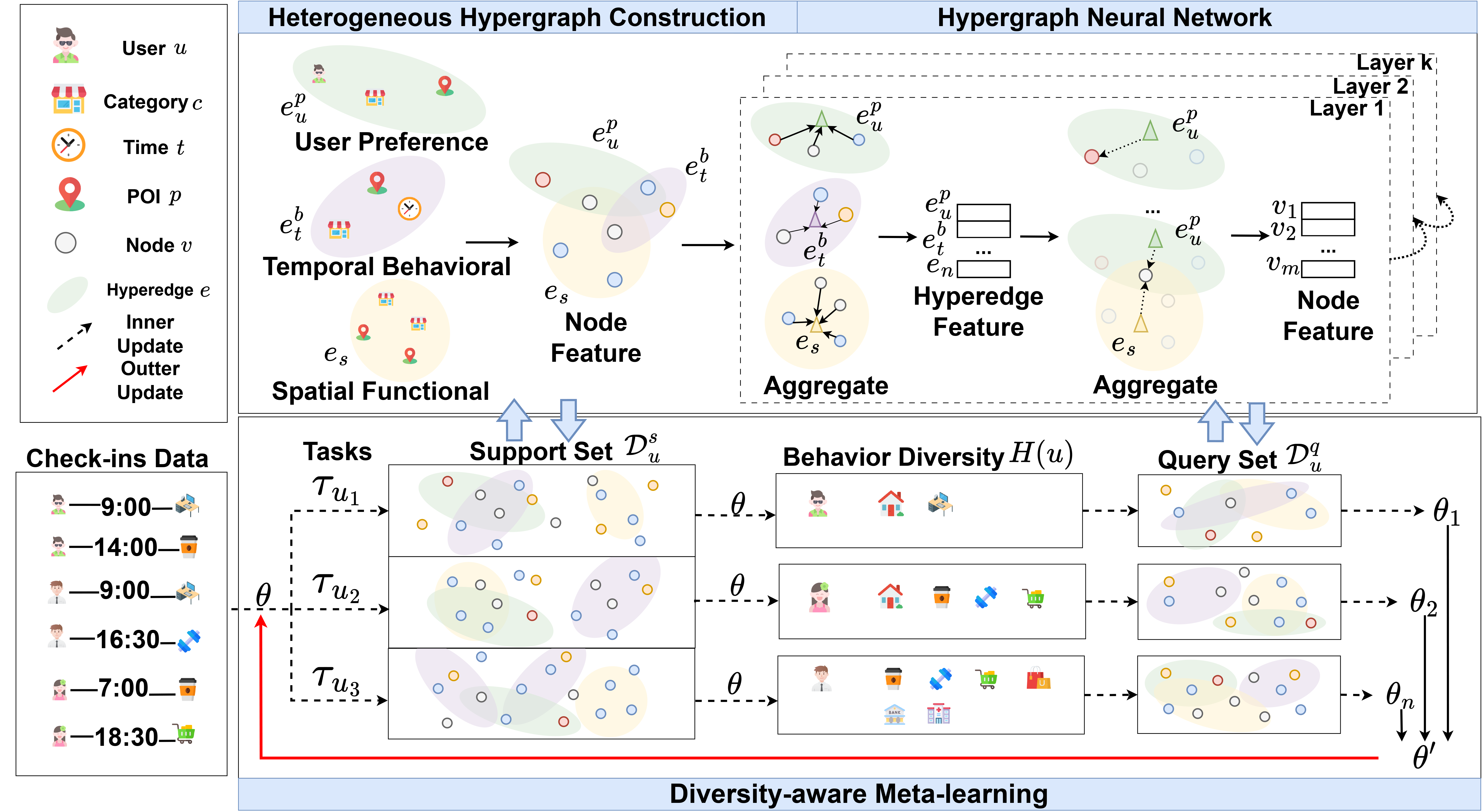} 
\caption{The overall framework of our proposed HyperMAN.}
\label{Fig:model} 
\vspace{-0.15in}
\end{figure*}

\smallskip\noindent\textbf{Meta-learning based Next POI Recommendation.}
Many existing methods employ meta-learning approaches to tackle the cold-start problem in next POI recommendation. For instance, MFNP~\cite{sun2021mfnp} combines region-specific user preferences by leveraging meta-learning for adaptive fusion, transferring knowledge from both individual and collective preferences. MetaKG~\cite{du2022metakg} employs collaborative-aware and knowledge-aware meta-learners to capture user preferences and knowledge associations, effectively addressing cold-start issues. However, these approaches do not sufficiently account for the diversity in behavioral patterns across different users. As a result, they may fail to capture the unique preferences and interactions of individual users, limiting their effectiveness in cold-start scenarios.

\textcolor{black}{\section{METHODOLOGY}}
\subsection{Problem Formulation}
\subsubsection{Next POI Recommendation}
Let $\mathcal{U}$, $\mathcal{P}$, and $\mathcal{C}$ denote the sets of users, POIs, and POI categories respectively. A check-in record is represented as $q = \langle u, p, c, g, t \rangle \in \mathcal{Q}$, where $u \in \mathcal{U}$, $p \in \mathcal{P}$, $c \in \mathcal{C}$, $g$ represents the geographical coordinates, and $t$ is the timestamp. Each trajectory $s = \{q_1, q_2, ..., q_n\} \in \mathcal{S}$ contains temporally ordered check-ins.

\subsubsection{Meta-learning Framework}
We formulate next POI recommendation as a meta-learning problem where each user's recommendation requirement is treated as a unique task $\tau_u$. 
For each $\tau_u$, we construct support set $\mathcal{D}^s_u$ and query set $\mathcal{D}^q_u$ from different periods of user's historical trajectories, thus each instance is a trajectory-POI pair $(s, p_{next})$. 
$\mathcal{L}(\theta, \mathcal{D})$ is the loss function that evaluates the model parameters $\theta$ on dataset $\mathcal{D} \in \mathcal{D}^s_u \cup \mathcal{D}^q_u$, 
and the learning iteration consists of an inner loop and an outer loop.

\noindent\textbf{Inner Loop:} For sampled tasks $\tau_u$, we construct support set $\mathcal{D}^s_u$ and query set $\mathcal{D}^q_u$ for each user. Then we perform inner update of $\theta$ by:
\begin{equation}
\label{eq:inner}
\theta' = \theta - \alpha\nabla_\theta\mathcal{L}(\theta, \mathcal{D}^s_u),
\end{equation}
where $\mathcal{L}$ is the cross-entropy loss measuring POI recommendation accuracy; $\alpha$ is the inner learning rate, and $\theta'$ is the inner updated parameters for each user.\\
\noindent\textbf{Outer Loop:} Using the inner updated parameters $\theta'$, we update the initialization $\theta$ by:
\begin{equation}
\label{eq:outer}
\theta = \theta - \beta\nabla_\theta \sum_{u \in \mathcal{U}} \mathcal{L}(\theta', \mathcal{D}^q_u),
\end{equation}
where $\beta$ is the outer loop learning rate.

\subsection{Heterogeneous Hypergraph Construction}

\noindent\textbf{Temporal Behavioral Hyperedge.} This hyperedge captures when a user visits a POI of specific category at certain time. A temporal behavioral hyperedge is defined as $e^b_t = \langle p, c, t \rangle$ where $p \in \mathcal{V}_p$ is a POI, $c \in \mathcal{V}_c$ is a category, and $t \in \mathcal{V}_t$ is a time slot. The set of temporal behavioral hyperedges is denoted as $E^b$. Let $v$ be any node in $\{p, c, t\}$. The corresponding incidence matrix is defined as:

\begin{equation}
\label{eq:temporal_incidence}
H_b(v,e_t^b) = \begin{cases}
1, & \text{if } v \in \{p, c, t\} \\
0, & \text{otherwise}
\end{cases}
\end{equation}

\noindent\textbf{Spatial Functional Hyperedge.} This hyperedge connects nearby POIs that share the same functional category. A spatial functional hyperedge is defined as $e^s = \langle p_i, p_j, c \rangle$ where POIs $p_i, p_j \in \mathcal{V}_p$ are within distance threshold $\delta$ and share the same category $c \in \mathcal{V}_c$. The set of spatial functional hyperedges is denoted as $E^s$. Let $v$ be any node in $\{p_i, p_j, c\}$. The corresponding incidence matrix is defined as:

\begin{equation}
\label{eq:spatial_incidence}
H_s(v,e^s) = \begin{cases}
1, & \text{if } v \in \{p_i, p_j, c\} \\
   &  \text{and } dist(p_i,p_j) \leq \delta \\
0, & \text{otherwise}
\end{cases}
\end{equation}

\noindent\textbf{User Preference Hyperedge.} This hyperedge groups a user's visited POIs of the same category to model long-term preferences. A user preference hyperedge is defined as $e^p_u = \langle u, p_1,\ldots,p_n, c \rangle$ where $u \in \mathcal{V}_u$ is a user, $p_1,\ldots,p_n \in \mathcal{V}_p$ are the POIs visited by user $u$, and $c \in \mathcal{V}_c$ is their corresponding category. The set of user preference hyperedges is denoted as $E^p$. Let $v$ be any node in $\{u, p_1,\ldots,p_n, c\}$. The corresponding incidence matrix is defined as:

\begin{equation}
\label{eq:preference_incidence}
H_p(v,e_u^p) = \begin{cases}
1, & \text{if } v \in \{u, p_1,\ldots,p_n, c\} \\
   & \text{ and } \{p_1,\ldots,p_n\} \subseteq \mathcal{Q}_u \\
0, & \text{otherwise}
\end{cases}
\end{equation}

Hence, the heterogeneous check-in hypergraph is defined as $G = (\mathcal{V}, E^b, E^s, E^p)$ which captures temporal patterns, spatial relationships, and user preferences simultaneously through different types of hyperedges.

\subsection{Hypergraph Neural Network}
For each node type $a \in \mathcal{A} = \{p, u, c, t\}$, we maintain a feature matrix $\mathbf{H}_a^{(l)} \in \mathbb{R}^{|V_a| \times d_a}$ at layer $l$, where $l = 1,\ldots,L$, $L$ is the number of layers, and $d_a$ is the feature dimension for type $a$. Let $\mathbf{D}_{v,r}$ denote the diagonal node degree matrix and $\mathbf{D}_{e,r}$ denote the diagonal hyperedge degree matrix for relation type $r \in \{b,s,p\}$. The message passing process for each type of hyperedge is defined as follows:

For temporal behavioral hyperedges:
\begin{equation}
\mathbf{Z}_b^{(l)} = \mathbf{D}_{v,b}^{-1/2} \mathbf{H}_b \mathbf{D}_{e,b}^{-1} \mathbf{H}_b^\top \mathbf{D}_{v,b}^{-1/2} \mathbf{H}_b^{(l)},
\end{equation}
where $\mathbf{H}_b$ is the incidence matrix defined in Eq.~\ref{eq:temporal_incidence}, $\mathbf{D}_{v,b}$ is the diagonal node degree matrix for temporal behavioral hyperedges, and $\mathbf{D}_{e,b}$ is the diagonal hyperedge degree matrix.

For spatial functional hyperedges:
\begin{equation}
\mathbf{Z}_s^{(l)} = \mathbf{D}_{v,s}^{-1/2} \mathbf{H}_s \mathbf{D}_{e,s}^{-1} \mathbf{H}_s^\top \mathbf{D}_{v,s}^{-1/2} \mathbf{H}_s^{(l)},
\end{equation}
where $\mathbf{H}_s$ is the incidence matrix defined in Eq.~\ref{eq:spatial_incidence}, $\mathbf{D}_{v,s}$ denotes the node degree matrix for spatial functional hyperedges, and $\mathbf{D}_{e,s}$ is the hyperedge degree matrix.

For user preference hyperedges:
\begin{equation}
\mathbf{Z}_p^{(l)} = \mathbf{D}_{v,p}^{-1/2} \mathbf{H}_p \mathbf{D}_{e,p}^{-1} \mathbf{H}_p^\top \mathbf{D}_{v,p}^{-1/2} \mathbf{H}_p^{(l)},
\end{equation}
where $\mathbf{H}_p$ is the incidence matrix defined in Eq.~\ref{eq:preference_incidence}, $\mathbf{D}_{v,p}$ represents the node degree matrix for user preference hyperedges, and $\mathbf{D}_{e,p}$ is the hyperedge degree matrix.

The aggregated message for node type $a$ is computed as:
\begin{equation}
\mathbf{Z}_a^{(l)} = \mathbf{W}_b \mathbf{Z}_b^{(l)} + \mathbf{W}_s \mathbf{Z}_s^{(l)} + \mathbf{W}_p \mathbf{Z}_p^{(l)},
\end{equation}
where $\mathbf{W}_b, \mathbf{W}_s, \mathbf{W}_p \in \mathbb{R}^{d_a \times d_a}$ are learnable weight matrices that balance the importance of different hyperedge types.

\begin{algorithm}[t]
\caption{HyperMAN Learning Process}
\label{alg:hyperman}
\footnotesize
\SetAlgoLined
\SetKwInOut{Input}{Input}
\SetKwInOut{Output}{Output}
\Input{Check-in records $\mathcal{Q}$, user set $\mathcal{U}$, POI set $\mathcal{P}$, category set $\mathcal{C}$}
\Output{Trained HyperMAN model parameters $\theta$}
\BlankLine

/* Heterogeneous Hypergraph Construction */ \;
Initialize nodes sets $\mathcal{V}_p$, $\mathcal{V}_u$, $\mathcal{V}_c$, $\mathcal{V}_t$ for POIs, users, categories, and time\;
Construct hyperedges $e^b_t$, $e^s$, $e^p_u$  using Eq.(3-6)\;
/* Hypergraph Neural Network */ \;
Initialize feature matrices $\{\mathbf{H}_a^{(0)}\}_{a \in \mathcal{A}}$ for each node type\;
Initialize type mapping matrices $\{\mathbf{M}_{r,a}\}$ and transformation parameters $\mathbf{\theta}$\;
\For{$l = 0$ \KwTo $L-1$}{
    \For{each node type $a \in \mathcal{A}$}{
        Update node representations $\mathbf{H}_a^{(l+1)}$ using Eq.(11)\;
    }
}
Obtain final node representations $\{\mathbf{H}_a^{(L)}\}_{a \in \mathcal{A}}$\;

/* Diversity-aware Meta-learning */ \;
\For{each user $u \in \mathcal{U}$}{
    Calculate behavior entropy $H(u)$ using Eq.(13)\;
    Compute adaptive learning rate $\alpha_u$ using Eq.(14)\;
    Sample support set $\mathcal{D}^s_u$ and query set $\mathcal{D}^q_u$\;
    Update user-specific parameters: $\theta'_u = \theta - \alpha_u\nabla_\theta\mathcal{L}(\mathcal{D}^s_u; \theta)$\;
    Update global parameters: $\theta \leftarrow \theta - \beta\nabla_\theta\mathcal{L}(\mathcal{D}^q_u; \theta'_u)$\;
}
\Return{$\theta$}
\end{algorithm}

The node feature update for the next layer is:
\begin{equation}
\mathbf{H}_a^{(l+1)} = \sigma(\mathbf{Z}_a^{(l)} \mathbf{M}_a \mathbf{H}_a^{(l)} \mathbf{\Theta}_a),
\end{equation}
where $\mathbf{M}_a \in \mathbb{R}^{d_a \times d_a}$ is a type-specific transformation matrix, $\mathbf{\Theta}_a \in \mathbb{R}^{d_a \times d_a}$ are learnable parameters, and $\sigma(\cdot)$ denotes the activation function.

To model sequential check-ins, we compute the trajectory representation via attention:
\begin{equation}
\label{eq:attention}
\alpha_i = \text{softmax}(\mathbf{W}_a[\mathbf{h}_{p_i}^{(L)}; \mathbf{h}_{c_i}^{(L)}; \mathbf{h}_{t_i}^{(L)}]),
\end{equation}
where $\mathbf{W}_a \in \mathbb{R}^{1 \times 3d}$ is a learnable attention weight matrix, and $[\cdot;\cdot;\cdot]$ denotes the concatenation operation.

\begin{equation}
\label{eq:trajectory}
\mathbf{z}_s = \sum_{i=1}^n \alpha_i[\mathbf{h}_{p_i}^{(L)}; \mathbf{h}_{c_i}^{(L)}; \mathbf{h}_{t_i}^{(L)}],
\end{equation}
where $\mathbf{h}_{p_i}^{(L)}$, $\mathbf{h}_{c_i}^{(L)}$, $\mathbf{h}_{t_i}^{(L)} \in \mathbb{R}^d$ are the final layer representations for POI, category, and time nodes respectively, and $i$ is the length of the trajectory.

\begin{table*}[t]
\footnotesize
\renewcommand{\arraystretch}{1}
\centering
\addtolength{\tabcolsep}{1pt}
\caption{Comparison results between HyperMAN and baselines on four datasets, 
where `R' stands to `Recall' and `N' stands `NDCG'; the best results are highlighted in bold, while the runner-up is underlined; the column `Improve' shows the improvements made by HyperMan compared to the runner-up.}\label{tab2}
\begin{tabular}{cl|cc|cc|cc|cc|r}
\toprule
\multicolumn{2}{c|}{\multirow{2}{*}{}} & \multicolumn{2}{c|}{Traditional}     & \multicolumn{2}{c|}{RNN-based}     & \multicolumn{2}{c|}{GNN-based}       & \multicolumn{2}{c|}{Meta Learning}     & \multirow{2}{*}{\textit{Improve}} \\
\multicolumn{2}{c|}{}                  & MostPop & \multicolumn{1}{c|}{BPRMF} & GRU    & \multicolumn{1}{c|}{STAN} & SGRec  & \multicolumn{1}{c|}{STHGCN} & MFNP         & \multicolumn{1}{c|}{HyperMAN} &                          \\ \midrule
\multirow{4}{*}{\rotatebox[origin=c]{90}{\textbf{NYC}}}       & R@5       & 0.0184  & 0.0366                     & 0.0734 & 0.1625                    & 0.3016 & \underline{0.3704}          & 0.3308       & \textbf{0.3957}               & 6.8\%                    \\
                           & N@5       & 0.0125  & 0.0339                     & 0.0392 & 0.1125                    & 0.2904 & \underline{0.3102}          & 0.2390       & \textbf{0.3534}               & 13.9\%                   \\
                           & R@10      & 0.2591  & 0.0556                     & 0.1624 & 0.2355                    & 0.3331 & \underline{0.3981}          & 0.3998       & \textbf{0.4262}               & 7.1\%                    \\
                           & N@10      & 0.2298  & 0.0294                     & 0.1467 & 0.2176                    & 0.2439 & \underline{0.2848}          & 0.2608       & \textbf{0.3164}               & 11.1\%                   \\ \hline 
\multirow{4}{*}{\rotatebox[origin=c]{90}{\textbf{PHO}}}       & R@5       & 0.0193  & 0.0415                     & 0.12   & 0.2535                    & 0.3368 & \underline{0.3801}          & 0.3421       & \textbf{0.4298}               & 13.0\%                   \\
                           & N@5       & 0.0151  & 0.0287                     & 0.0842 & 0.2422                    & 0.3076 & \underline{0.3539}          & 0.2282       & \textbf{0.3932}               & 11.0\%                   \\
                           & R@10      & 0.0277  & 0.0583                     & 0.1559 & 0.2621                    & 0.3963 & 0.3951                      & \underline{0.4022} & \textbf{0.4525}               & 12.5\%                   \\
                           & N@10      & 0.0231  & 0.0311                     & 0.1379 & 0.2489                    & 0.2821 & 0.2763                      & \underline{0.3487} & \textbf{0.3834}               & 9.9\%                    \\ \hline
\multirow{4}{*}{\rotatebox[origin=c]{90}{\textbf{SIN}}}       & R@5       & 0.0110  & 0.0394                     & 0.0915 & 0.1900                    & 0.2515 & \underline{0.3075}          & 0.2935       & \textbf{0.3253}               & 5.7\%                    \\
                           & N@5       & 0.0117  & 0.0221                     & 0.0691 & 0.1243                    & 0.2401 & 0.2235                      & \underline{0.2439} & \textbf{0.2679}               & 9.8\%                    \\
                           & R@10      & 0.0379  & 0.0542                     & 0.1416 & 0.2452                    & 0.2949 & \underline{0.3334}          & 0.3018       & \textbf{0.3575}               & 7.2\%                    \\
                           & N@10      & 0.1878  & 0.0361                     & 0.1022 & 0.2024                    & 0.2348 & \underline{0.2781}          & 0.2526       & \textbf{0.2910}               & 4.6\%                    \\ \hline
\multirow{4}{*}{\rotatebox[origin=c]{90}{\textbf{CAL}}}       & R@5       & 0.0571  & 0.0744                     & 0.1836 & 0.2851                    & 0.3324 & \underline{0.4093}          & 0.3535       & \textbf{0.4702}               & 14.8\%                   \\
                           & N@5       & 0.0363  & 0.0445                     & 0.1325 & 0.2392                    & 0.2314 & \underline{0.3683}          & 0.3056       & \textbf{0.4206}               & 14.2\%                   \\
                           & R@10      & 0.0913  & 0.0118                     & 0.2236 & 0.3296                    & 0.3609 & \underline{0.4445}          & 0.3793       & \textbf{0.5261}               & 18.3\%                   \\
                           & N@10      & 0.0417  & 0.0859                     & 0.1705 & 0.2403                    & 0.2831 & \underline{0.3907}          & 0.3031       & \textbf{0.4429}               & 13.3\%                   \\ \bottomrule
\end{tabular}

\end{table*}

\subsection{Diversity-Aware Meta-Learning}
Let $\mathcal{D}_u = \{(s_i, p_{i+1})\}_{i=1}^{N_u}$ denote the sequence of check-ins for user $u$, where $s_i$ represents the historical trajectory and $p_{i+1}$ is the next POI to visit. We split $\mathcal{D}_u$ into support set $\mathcal{D}^s_u$ and query set $\mathcal{D}^q_u$ for meta-learning.

For each user $u$, we measure behavior diversity through category-based visit entropy:
\begin{equation}
\label{eq:entropy1}
H(u) = -\sum_{c \in \mathcal{C}} p(c|u) \log p(c|u),
\end{equation}
where $p(c|u)$ represents the proportion of check-in records with POI category $c$ among all check-in records of user $u$. 
Based on this measure, Eq.(1) is reformulated as: 
\begin{equation}
\label{eq:adaptive_rate1}
\theta'_u = \theta - \alpha_u \nabla_\theta \mathcal{L}(\theta, \mathcal{D}^s_u),
\end{equation}

\begin{equation}
\label{eq:adaptive_rate2}
\alpha_u = \alpha_0 \cdot \text{sigmoid}(\beta H(u)),
\end{equation}
where $\alpha_0$ is the base learning rate and $\beta$ controls the entropy sensitivity.

Accordingly, the probability distribution over all candidate next POIs is determined using the softmax function, given by,
\begin{equation}
\label{eq:prediction}
P(p|s) = \text{softmax}(\mathbf{W}[\mathbf{z}_s; \mathbf{h}_p^{(L)}; \mathbf{h}_c^{(L)}] + \mathbf{b}),
\end{equation}
where $\mathbf{W} \in \mathbb{R}^{|\mathcal{P}| \times 3d}$ is the learnable weight matrix and $\mathbf{b} \in \mathbb{R}^{|\mathcal{P}|}$ is the bias term.
\begin{figure*}[t]
\centering 
\includegraphics[width=0.77\textwidth]{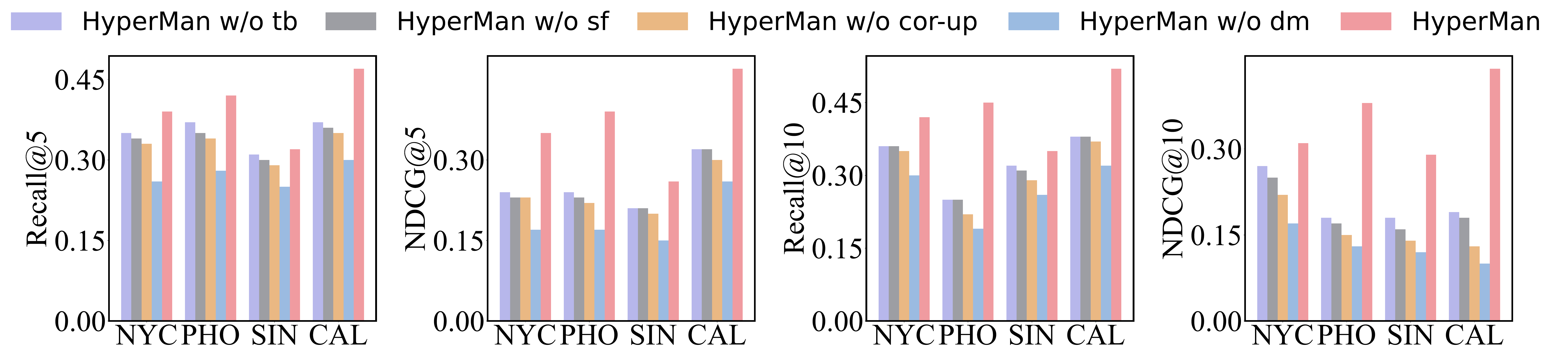} 
\vspace{0.01in}
\caption{Performance comparison for variants of HyperMAN on the four datasets.}
\label{Fig:ablation} 
\vspace{-0.15in}
\end{figure*}

\begin{figure}[t]
\centering
\begin{minipage}[t]{0.36\linewidth}
\centering
\includegraphics[width=\textwidth]{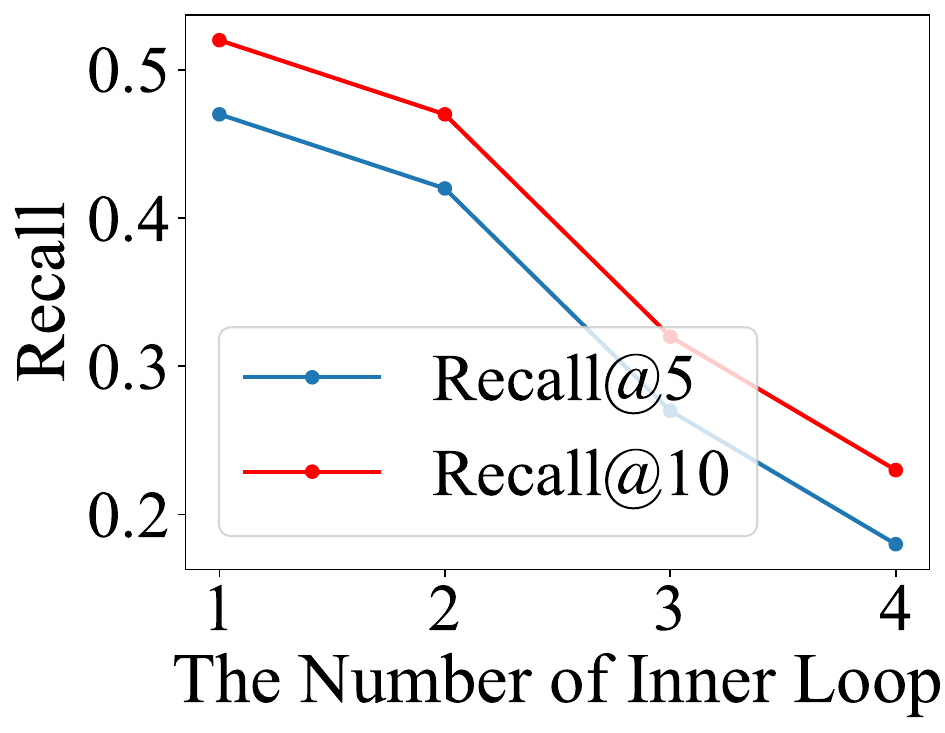}
\end{minipage}
\hspace{0.05\linewidth}
\begin{minipage}[t]{0.36\linewidth}
\centering
\includegraphics[width=\textwidth]{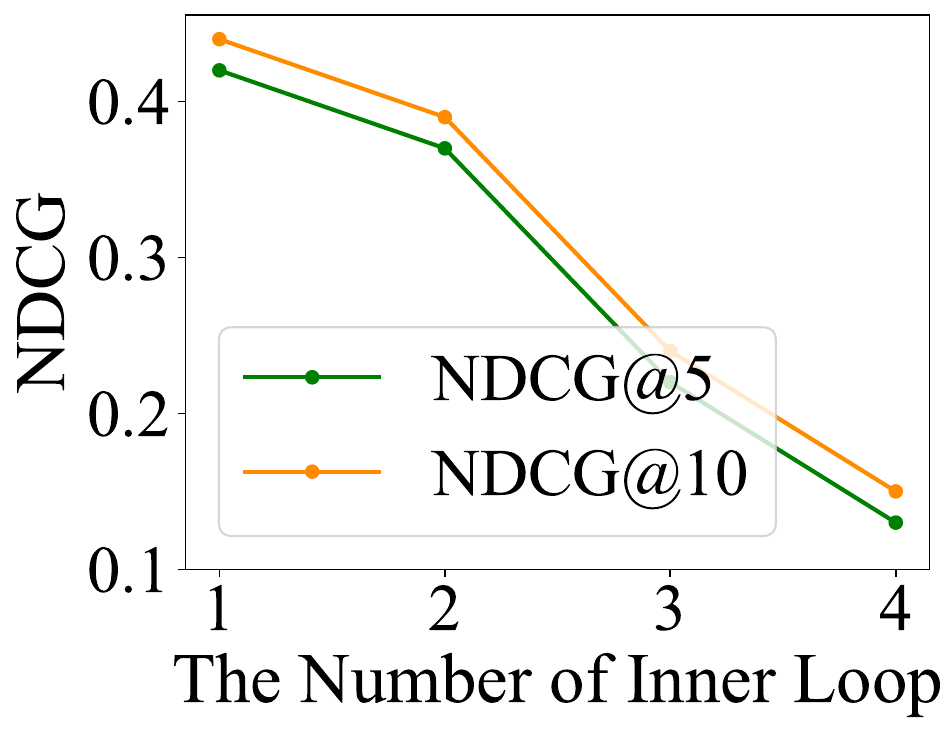}
\end{minipage}
\caption{Sensitivity analysis for Inner Loop results.}\label{fig:sensitivity}
\vspace{-0.15in}
\end{figure}

\begin{figure}[t]
\centering
\begin{minipage}[t]{0.36\linewidth}
\centering
\includegraphics[width=\textwidth]{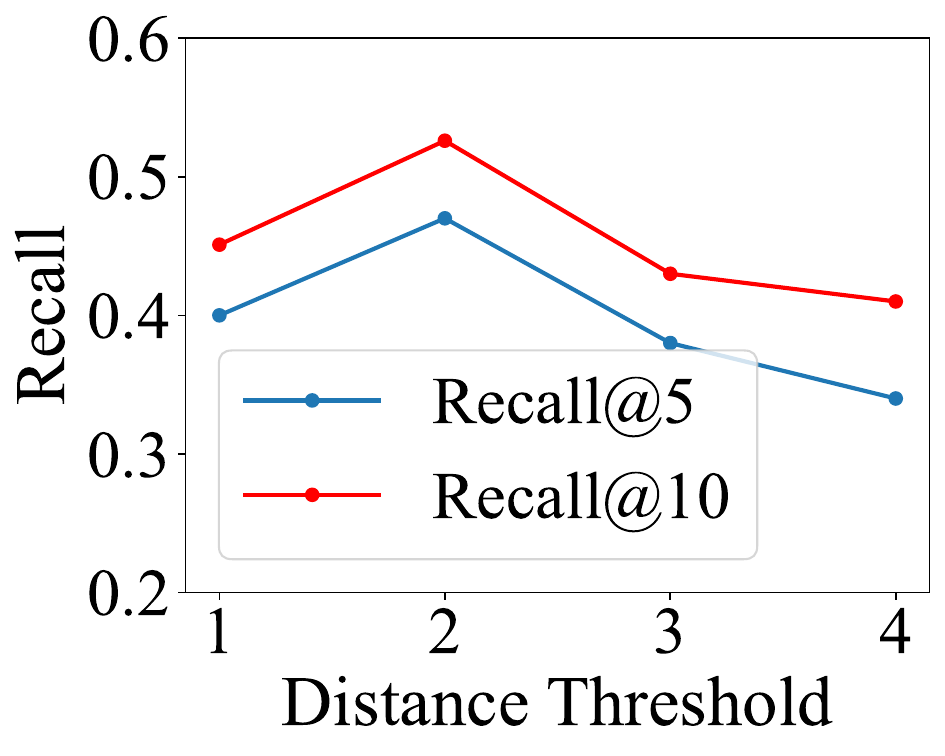}
\end{minipage}
\hspace{0.05\linewidth}
\begin{minipage}[t]{0.36\linewidth}
\centering
\includegraphics[width=\textwidth]{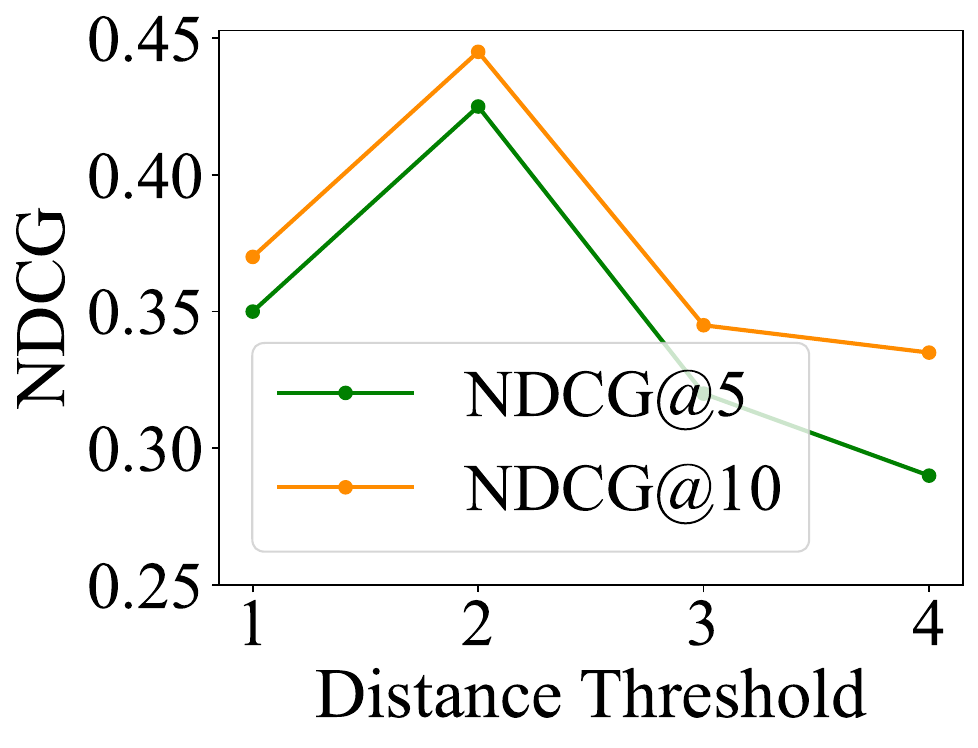}
\end{minipage}
\caption{Impact of distance threshold on performance metrics.}\label{fig:distance}
\vspace{-0.2in}
\end{figure}

\section{Experiments and Results}

We conduct experiments\footnote{Our data and codes are available at: \url{https://github.com/ICME-2025/HyperMAN}.} to answer three research questions: \textbf{(RQ1)} How does HyperMAN's performance compare to existing state-of-the-art approaches in next POI recommendation? \textbf{(RQ2)} What is the impact of different components on the performance of HyperMAN? \textbf{(RQ3)} How do hyper-parameter settings influence HyperMAN? 

\subsection{Experimental Settings}
\noindent\textbf{Datasets and Metrics.} We conduct experiments on the widely used public dataset Foursquare and select four cities (i.e., New York, Singapore, Phoenix, and Calgary) for evaluation. 
For each city, we split each user's check-ins into new successive sequences within a day~\cite{zhangnext,wang2023meta}. 
Two standard evaluation metrics, $Recall@K$ and $NDCG@K$, are adopted following~\cite{lai2024disentangled}. 
The former quantifies the proportion of user-preferred items successfully identified within the top-K recommendations, the latter evaluates the ranking quality of relevant items in the top-K results. Each experiment is conducted ten times to ensure reliability, and the average $Recall@K$ and $NDCG@K$ values are reported accordingly. 

\noindent\textbf{Baselines.} 
We compare our proposed HyperMAN with several state-of-the-art baselines:
(1) \texttt{MostPop}, a method that predicts the next POI based on the frequency of a user's visits; (2) \texttt{BPRMF}~\cite{rendle2009bpr}, a matrix factorization method refined through Bayesian personalized ranking optimization; (3) \texttt{GRU}~\cite{cho2014learning}, an RNN variant that regulates information flow through two gating mechanisms; (4) \texttt{STAN}~\cite{luo2021stan}, a self-attention model designed to explicitly capture spatio-temporal dynamics in a user's check-in sequence; (5) \texttt{SGRec}~\cite{kang2018self}, a GNN-based approach modeling collaborative interactions among neighboring nodes; (6) \texttt{STHGCN}~\cite{yan2023spatio}, a hypergraph model that combines intricate high-order information and global collaborative dependencies across trajectories; and (7) \texttt{MFNP}~\cite{sun2021mfnp}, a meta-learning model that leverages region-level user preferences by adaptively fusing personal and crowd knowledge.

\subsection{Performance Comparison (RQ1)} The results are presented in Table~\ref{tab2}. Across the four datasets, traditional methods (MostPop, BPRMF) generally underperform compared to RNN-based methods (GRU, STAN), highlighting the effectiveness of sequential neural networks in producing more accurate recommendations. Among RNN-based methods, STAN outperforms GRU by utilizing relative spatio-temporal data via self-attention layers. GNN-based methods (SGRec, STHGCN) demonstrate superior performance over RNN-based methods, underscoring the capability of GNNs in modeling complex dependencies. The superior performance of STHGCN over SGRec further reflects the ability of hypergraphs to capture high-order relationships effectively. MFNP, as a representative meta-learning-based method, achieves marginally better results than STHGCN on specific metrics but generally outperforms SGRec, indicating that standalone meta-learning methods still face challenges in capturing high-order information. Our proposed HyperMAN, which integrates meta-learning with hypergraph modeling, consistently achieves the best performance across all datasets, with an average improvement of 10.67\% and 10.97\% in Recall and NDCG, respectively. These results confirm the advantages of heterogeneous hypergraph structures and diversity-aware meta-learning in enhancing learning effectiveness.

\subsection{Ablation Study (RQ2)} 
We investigate the impacts of different components in HyperMAN: 
(1) HyperMan$_{w/o \ tb}$ removes the temporal behavioral hyperedge from edge construction; (2) HyperMan$_{w/o \ sf}$ removes the spatial functional hyperedge from edge construction; (3) HyperMan$_{w/o \ up}$ removes the user preference hyperedge from edge construction; (4) HyperMan$_{w/o \ dm}$ removes the diversity-aware meta-learning, but only retains the MAML for meta-training. 
From Fig.~\ref{Fig:ablation}, we observe that HyperMan${w/o \ up}$ outperforms both HyperMan${w/o \ tb}$ and HyperMan$_{w/o \ sf}$, indicating that the user preference hyperedge plays a more significant role than the temporal behavioral hyperedge and spatial functional hyperedge. Besides, HyperMan$_{w/o \ dm}$ performs worse than HyperMan, confirming the advantage of diversity-aware meta-learning.

\subsection{Sensitivity Analysis (RQ3).} We study the influence of two key hyperparameters, i.e., the number of inner-update steps in Eq.~(\ref{eq:inner}) and the distance threshold used in constructing the spatial functional hyperedge on NYC dataset. Similar patterns are observed across the other three datasets. 
Fig.~\ref{fig:sensitivity} illustrates the model performance to the number of inner-update steps, which suggests that a single update step is sufficient to achieve optimal recommendation performance, while also enhancing model efficiency. Fig.~\ref{fig:distance} explores the effect of the distance threshold on the construction of spatial functional hyperedges. 
As the distance threshold increases, the performance initially improves before slightly declining, indicating that a larger range helps capture a broader spectrum of potential user interests. However, excessive expansion of the range may compromise the model's accuracy.
\section{Conclusion}
In this paper, we propose the Hypergraph-enhanced Meta-learning Adaptive Network (HyperMAN) for next POI recommendation, addressing the cold-start problem. Three types of heterogeneous hyperedges are constructed to capture complex high-order relationships: temporal behavioral hyperedges, spatial functional hyperedges, and user preference hyperedges. A diversity-aware meta-learning training approach is designed to account for user behavioral diversity. Experimental results on real-world datasets demonstrate the effectiveness of HyperMAN in improving recommendation accuracy.

%
%
%
\bibliographystyle{splncs04}
\bibliography{mybibliography}

\begin{thebibliography}{10}
\providecommand{\url}[1]{\texttt{#1}}
\providecommand{\urlprefix}{URL }
\providecommand{\doi}[1]{https://doi.org/#1}

\bibitem{cai2023user}
Cai, D., et~al.: User cold-start recommendation via inductive heterogeneous graph neural network. TOIS  \textbf{41}(3),  1--27 (2023)

\bibitem{cheng2013you}
Cheng, C., et~al.: Where you like to go next: Successive point-of-interest recommendation. In: IJCAI. pp. 2605--2611 (2013)

\bibitem{cho2014learning}
Cho, K., et~al.: Learning phrase representations using rnn encoder-decoder for statistical machine translation. arXiv:1406.1078  (2014)

\bibitem{du2022metakg}
Du, Y., et~al.: Metakg: Meta-learning on knowledge graph for cold-start recommendation. TKDE  \textbf{35}(10),  9850--9863 (2022)

\bibitem{halder2021transformer}
Halder, S., et~al.: Transformer-based multi-task learning for queuing time aware next poi recommendation. In: PAKDD. pp. 510--523 (2021)

\bibitem{ju2022kernel}
Ju, W., et~al.: Kernel-based substructure exploration for next poi recommendation. In: ICDM. pp. 221--230 (2022)

\bibitem{kang2018self}
Kang, W.C., et~al.: Self-attentive sequential recommendation. In: ICDM. pp. 197--206 (2018)

\bibitem{lai2024disentangled}
Lai, Y., et~al.: Disentangled contrastive hypergraph learning for next poi recommendation. In: SIGIR. pp. 1452--1462 (2024)

\bibitem{lim2020stp}
Lim, N., et~al.: Stp-udgat: Spatial-temporal-preference user dimensional graph attention network for next poi recommendation. In: CIKM. pp. 845--854 (2020)

\bibitem{liu2023mandari}
Liu, X., et~al.: Mandari: Multi-modal temporal knowledge graph-aware sub-graph embedding for next-poi recommendation. In: ICME. pp. 1529--1534 (2023)

\bibitem{luo2021stan}
Luo, Y., et~al.: Stan: Spatio-temporal attention network for next location recommendation. In: WWW. pp. 2177--2185 (2021)

\bibitem{rendle2009bpr}
Rendle, S., et~al.: Bpr: Bayesian personalized ranking from implicit feedback. In: UAI. pp. 452--461 (2009)

\bibitem{sun2021mfnp}
Sun, H., et~al.: Mfnp: A meta-optimized model for few-shot next poi recommendation. In: IJCAI. pp. 3017--3023 (2021)

\bibitem{tan2024heterogeneous}
Tan, Y., et~al.: Heterogeneous hypergraph structure learning for multimedia recommendation. In: ICME. pp.~1--6 (2024)

\bibitem{wang2023meta}
Wang, J., et~al.: Meta-learning enhanced next poi recommendation by leveraging check-ins from auxiliary cities. In: PAKDD. pp. 322--334 (2023)

\bibitem{wang2023eedn}
Wang, X., et~al.: Eedn: Enhanced encoder-decoder network with local and global context learning for poi recommendation. In: SIGIR. pp. 383--392 (2023)

\bibitem{wang2023adaptive}
Wang, Z., et~al.: Adaptive graph representation learning for next poi recommendation. In: SIGIR. pp. 393--402 (2023)

\bibitem{yan2023spatio}
Yan, X., et~al.: Spatio-temporal hypergraph learning for next poi recommendation. In: SIGIR. pp. 403--412 (2023)

\bibitem{zhangnext}
Zhang, L., et~al.: Next point-of-interest recommendation with inferring multi-step future preferences. In: IJCAI. pp. 3751--3757 (2022)

\bibitem{zhang2024dshgt}
Zhang, T., et~al.: Dshgt: Dual-supervisors heterogeneous graph transformer—a pioneer study of using heterogeneous graph learning for detecting software vulnerabilities. TOSEM  \textbf{33}(8),  1--31 (2024)

\bibitem{zhang2025learning}
Zhang, T., et~al.: Learning from heterogeneity: A dynamic learning framework for hypergraphs. TAI  (2025)

\bibitem{zhao2018personalized}
Zhao, S., et~al.: Personalized sequential check-in prediction: Beyond geographical and temporal contexts. In: ICME. pp.~1--6 (2018)

\end{thebibliography}

\end{document}